\documentclass[a4paper,11pt]{article}
\usepackage{jheppub} 
\usepackage{siunitx}
\usepackage[utf8]{inputenc}
\usepackage{amsmath}
\usepackage{amsfonts}
\usepackage{amssymb}
\usepackage[T1]{fontenc}
\usepackage{siunitx}
\usepackage{float}
\usepackage[english]{babel}
\usepackage[caption=false]{subfig}

\usepackage{slashed}
\usepackage{graphicx}
\usepackage{datetime}

\title{Constraints on Light Leptophilic Dark Matter Mediators from Decay
  Experiments}
\author[]{Gerrit Bickendorf,}
\author[]{Manuel Drees}
\emailAdd{drees@th.physik.uni-bonn.de}
\emailAdd{bickendorf@th.physik.uni-bonn.de}
\affiliation[]{Bethe Center for Theoretical Physics and Physikalisches
  Institut, Universit\"{a}t Bonn, Nussallee 12, D-53115 Bonn, Germany}

\abstract{We study the influence of leptophilic dark matter
  interactions on decays of muons and ground state mesons in existing
  experiments. We consider a secluded dark sector exclusively
  interacting with leptons via either a (leptophilic) scalar or vector
  mediator. These interactions will therefore influence leptonic
  decays and deform the energy spectra. We first study the Michel
  decay of muons, $\mu^+\rightarrow e^+\nu_e \bar{\nu}_\mu$, which
  allow us to constrain the parameter space reasonably well. Secondly,
  the rare $\pi^\pm$, $K^\pm$, $D^\pm$ and $D_s^\pm$ decays to $e\nu$
  will be considered. Scalar mediators would remove the Standard Model
  helicity suppression, so that strong constraints can be derived. The
  resulting bounds on the couplings of the light mediators to
  electrons and muons still turn out to be somewhat weaker than those
  from searches at low--energy $e^+e^-$ colliders and the magnetic
  moment of the muon, respectively. Finally, we show that kaon and
  pion decays basically exclude a ``Co--SIMP'' scenario where a scalar
  dark matter particle has a dimension--5 coupling to electrons.}

\makeatletter
\gdef\@fpheader{}
\makeatother
\begin{document}
\def\gsim{\:\raisebox{-0.5ex}{$\stackrel{\textstyle>}{\sim}$}\:}
\def\lsim{\:\raisebox{-0.5ex}{$\stackrel{\textstyle<}{\sim}$}\:}
\maketitle
\raggedbottom

\section{Introduction}

Evidence for the existence of substantial non--baryonic mass in the
universe, in addition to the baryonic contribution from the known
Standard Model (SM) particles, has been piling up for decades
\cite{ParticleDataGroup:2020ssz}. Cosmological observations, from the
CMB at the largest scales, the structure of galaxy clusters and
gravitational lensing on intermediate scales, down to the rotation of
single galaxies, all essentially only probe the gravitational
interactions of this dark matter (DM), leaving the properties of the
constituents of DM largely obscure. We do know that DM should be
``cold'', i.e. non--relativistic well before the CMB
decoupled. Moreover, within the minimal cosmological framework, the
overall DM density can be determined accurately,
$\Omega_{\rm DM} h^2 =0.120\pm 0.001$ \cite{collaboration2018planck};
here $\Omega_{\rm DM}$ is the scaled DM mass density and $h$ is the
rescaled Hubble parameter.

In the absence of data pinning down the properties of DM, a wide range
of models has been proposed. In spite of intensive efforts no clear
signal for DM particles has yet been found in either direct or
indirect detection experiments \cite{ParticleDataGroup:2020ssz},
leading to severe constraints on many models
\cite{Roszkowski_2018}. This is true in particular for models with
weakly interacting massive particles (WIMPs), with masses very roughly
at the weak scale. This has led to increased interest in sub--GeV
masses \cite{Knapen_2017}, which for purely kinematic reasons are much
less constrained.

Here we consider models that couple the potential dark sector to the
Standard Model with light leptophilic mediators, i.e. mediators that
couple directly only to leptons. One class of models assumes a gauged
lepton--family number; after spontaneous symmetry breaking these
models contain a massive vector mediator, sometimes called a ``dark
photon'', that couples to some leptons of the Standard Model and in
the dark sector to Dirac dark matter \cite{Altmannshofer_2016,
  essig2013dark}. This lepton coupling introduces a kinetic mixing
term with the ordinary photon, resulting in a small coupling between
the dark photon and all electrically charged particles which can also
be used to constrain the model. Another class of models assumes a
scalar mediator, which again only couples directly to some or all
charged leptons. Since this breaks the $SU(2)$ gauge symmetry, models
of this kind can at most be an effective theory. Finally, we consider
the so--called ``Co--SIMP'' model \cite{Smirnov:2020zwf} containing a
light scalar DM particle with a non--renormalizable coupling to
electrons.

We present a novel approach to constraining the parameter space using
measurements of the decays of muons or ground--state flavored mesons
into final states containing an electron. The spectrum of electrons
produced in muon decays has been measured accurately; it agrees with
SM predictions, which allows us to put upper bounds on the coupling of
spin$-1$ mediators to electrons or muons. However, the resulting
bounds turn out to be more than one order of magnitude weaker than the
best constraint from $e^+e^-$ colliders. New spin$-0$ particles
coupling to electrons would remove the helicity suppression in charged
meson decays into $e \nu_e$ final states; the resulting bounds on the
renormalizable couplings of light scalar mediators are tighter than
those from muon decay, but still somewhat weaker than those from
$e^+e^-$ colliders. However, bounds from pion and kaon decays suffice
to exclude a thermal ``Co--SIMP'' for masses below $0.8$ MeV; in
the allowed mass range $\chi$ does not behave like a SIMP any more.

The remainder of this article is structured as follows: In section
\ref{se:LepDM} we present the leptophilic models considered here. In
sections \ref{se:MuBounds} and \ref{se:BoundsPI} we describe the
method of obtaining limits on the parameter space from the Michel
decays of muons and pseudoscalar meson decays respectively. Section
\ref{se:Results} presents our results and compares them to existing
bounds. Section \ref{se:Conclusion} finishes with some concluding
remarks.

\section{Models}
\label{se:LepDM}

The strongest bound on many DM models comes from ``direct'' search
experiments, which look for elastic scattering of ambient DM particles
off the nuclei in a detector \cite{ParticleDataGroup:2020ssz}.
Leptophilic dark matter models, where the dark matter particles
primarily couple to the Standard Model leptons, either directly or via
another ``mediator'' particle, avoid most of these bounds. In such
models the DM particles can interact with nucleons only via loop
diagrams.

One way of incorporating these ideas is a hidden sector that contains
only singlets under the Standard Model gauge group. However, the
simplest (thermal) DM production mechanism requires some coupling to
SM particles.  To this end one may introduce additional fields which
mediate interactions between both sectors. These are then fittingly
called portals.

\subsection{Vector Mediator}

The extension of the Standard Model by a new vector boson is well
motivated both from a bottom up as well as from a top down
perspective, e.g. from grand unified theories
\cite{delAguila:1988jz}. Here we consider scenarios where the gauge
group is extended by another $U(1)_D$ gauge group which is
spontaneously broken such that the associated particles become
massive. The new vector boson $A'$ is often called a dark photon.  By
assumption the Dark Matter particles $\chi$ are charged under
$U(1)_D$. In this article we are concerned with the production of a light
$A'$ which decays invisibly. The exact nature of $\chi$ therefore is not
relevant for us. Assuming it to be a Dirac fermion for simplicity's
sake, the resulting Lagrangian can the be written as
\begin{align} \label{eq:L}
  \mathcal{L}=& \mathcal{L}_\text{SM}-\frac{1}{4}F_{\mu\nu}'F'^{\mu\nu} +
                \frac{m_{A'}^2}{2}A_\mu'A'^\mu +\frac{\epsilon}{2}F_{\mu\nu}'
                F^{\mu\nu}-\sum_{l=e,\mu,\tau} e'_l\left(\bar{l}\gamma^\mu A'_\mu
                l + \bar{\nu}_l\gamma^\mu A'_\mu \nu_l\right)
  \\ \nonumber
              &+\bar{\chi}(i\slashed{\partial}-m_\chi)\chi -
                g_D\bar{\chi}\gamma^\mu A_\mu'\chi\,.
\end{align}
Here $F_{\mu\nu}$ and $F'_{\mu\nu}$ are the field strength tensors of
QED and of $U(1)_D$, respectively. The new coupling constants $g_D$
and $e_l'$ are free parameters of the model, and the term
$\propto \epsilon$ describes kinetic mixing between the new gauge
boson and the photon. In order to avoid anomalies, one may chose to
gauge any combination $X=yB-\sum x_iL_i$ of baryon number $B$ and
lepton family number $L_i$, with constraint $3y=x_e+x_\mu+x_\tau$
\cite{Altmannshofer:2014pba}. Popular choices include gauged $B-L$ or
$L_\mu-L_\tau$ \cite{Bhupal_Dev_2021, Escudero_2019}. While the former
is highly constrained by direct collider searches and other 5th force
experiments, the latter still exhibits rather weak constraints with a
region that is even favored by the $g_\mu -2$ anomaly
\cite{Fayet:2007ua, Pospelov:2008zw, Foldenauer_2019}.

\begin{figure}[H]
  \centering
    \includegraphics[page=1]{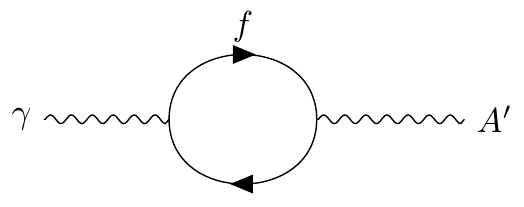}
    \caption{Kinetic mixing of $A'$ with the SM photon.}
\label{fg:KinMix}
\end{figure}

The term proportional to $\epsilon$ has been added to work with the
most general renormalizable gauge invariant Lagrangian.  Even though
this term might be absent at tree level, as is the case for some GUT
theories, it can be generated by loop contributions such as the
diagram shown in fig.~\ref{fg:KinMix}, where the fermion running in
the loop is charged both under the Standard Model and dark
$U(1)$-group. In the case at hand this leads to
\begin{equation} \label{eq:KinMix}
\epsilon =\sum_l \frac{ee_l'}{12\pi^2}\ln\left(\frac{m_l^2}{\mu^2}\right) \,.
\end{equation}
In anomaly--free theories the dependence on the renormalization scale
$\mu$ cancels in the sum. The term $\propto \epsilon$ in
eq.(\ref{eq:L}) leads to additional effective interactions of the form
\begin{equation} \label{eq:Leps}
\mathcal{L}\supset e\epsilon A'_\mu J_{\rm em}^\mu \,,
\end{equation}
where $J_{\rm em}^\mu = \sum_\psi Q_\psi \bar \psi \gamma^\mu \psi$ is
the electromagnetic current; hence every electrically charged particle
interacts with the dark photon as it is now millicharged under
$U(1)_D$ \cite{Rizzo_2019}. This can be used to put strong constraints
on the parameter space.

If $m_{A'}<2m_\chi$ the new gauge boson will mostly decay to the
leptons to which it couples directly:
\begin{equation} \label{eq:Adec}
  \Gamma(A' \rightarrow l \bar l) = \frac{e_l^{'2}}{12\pi} d_l
  m_{A'}\left(
    1+\frac{2m_l^2}{m_{A'}^2}\right)\sqrt{1-\frac{4m_l^2}{m_{A'}^2}}\,,
\end{equation}
where $d_l = 1$ for charged leptons while $d_l = 1/2$ for left--handed
neutrinos. If $m_{A'} > 2 m_{l^\pm}$ one may search for visible
$A' \rightarrow l^+l^-$ decays. Here we are instead interested in
scenarios with mostly invisible $A'$ decays, either because
$m_{A'} < 2 m_{l^\pm}$ for the relevant charged lepton, or because
$m_{A'}>2m_\chi$ and $g_D \gg e'_l$. $A' \rightarrow \chi \bar\chi$
decays are also described by eq.(\ref{eq:Adec}), with the obvious
replacements $e_l \rightarrow g_D, \ m_l \rightarrow m_\chi$ and
$d_l \rightarrow 1$.

Within a given cosmological scenario the $\chi$ relic density imposes
one constraint on the parameters of the model \cite{Knapen_2017,
  Foldenauer_2019}; even the case $e'_l = 0$, in which case $A'$
couples to SM particles only via eq.(\ref{eq:Leps}), can lead to the
correct relic density in minimal cosmology
\cite{Izaguirre:2014bca}. Here we implicitly assume that this
constraint is used to determine the DM mass $m_\chi$, which allows us
to vary the mass and couplings of $A'$ freely.

Additional motivation for direct interactions with the muon
specifically come from the anomalous magnetic moment of the muon,
$a_\mu = (g-2)_\mu/2$. The Standard Model prediction
\cite{Aoyama:2020ynm} differs from the experimental result
\cite{PhysRevLett.126.141801}:
$\Delta a_\mu= a_\mu(\text{Exp}) - a_\mu(\text{SM})= (251\pm59) \times
10^{-11}$. The additional $1-$loop contribution from a vector boson
coupling to muons is \cite{LEVEILLE197863, Kahn:2018cqs}:
\begin{equation} \label{eq:amu_A}
  \Delta a_\mu^{A'}= \frac{e_\mu'^2} {4\pi^2} \int_0^1 d z \frac{m_\mu^2z(1-z)^2}
  {m_\mu^2(1-z)^2+m_{A'}^2z}\,.
\end{equation}
For $m_{A'}\ll m_\mu$ this simplifies to
\begin{equation}
  \Delta a_\mu^{A'} \approx \frac{e_\mu'^2}{8 \pi^2}
  \approx 1.3\times 10^{-10}\left(\frac{e_\mu'}{10^{-4}}\right)^2\,.
\end{equation}

\subsection{Scalar mediator}

Another possibility is that a scalar particle mediates interactions
between the dark matter and the SM particles. Here we consider a real
scalar field $\phi$ with mass $m_\phi$ that is a singlet under the
Standard Model gauge group, and again a dark matter Dirac fermion $\chi$.
The Lagrangian is:
\begin{equation} \label{eq:L_scal}
  \mathcal{L}= \mathcal{L}_\text{SM} + \frac{1}{2}\partial_\mu \phi \partial^\mu
  \phi - \frac{m_{\phi}^2}{2}\phi^2 - \sum_{l=e,\mu,\tau} e'_l\bar{l} l \phi
 +\bar{\chi}(i\slashed{\partial}-m_\chi)\chi - g_D\bar{\chi}\chi\phi\, .
\end{equation}
This Lagrangian respects QED gauge invariance, but the interactions
of the scalar with the leptons break electroweak gauge invariance
explicitly. These couplings might originate from gauge invariant
(but non--renormalizable) dimension $5$ operators \cite{Batell:2016ove}:
\begin{equation} \label{eq:dim_5}
\frac{c_l}{\Lambda}\phi \bar{L}_i\Phi e_{iR}+h.c. \,,
\end{equation}
where $\Phi$ is the Standard Model Higgs field. Once $\Phi$ obtains
a vacuum expectation value $v$, the interactions in (\ref{eq:dim_5})
lead to the scalar couplings in eq.(\ref{eq:L_scal}), with
\begin{equation}
e'_l=\frac{c_l v}{\Lambda\sqrt{2}}\,.
\end{equation}
Another possibility is to have the light scalar $\phi$ mix with the
neutral component of a (second) scalar doublet, which can in principle
have renormalizable ${\cal O}(1)$ couplings to leptons.\footnote{Such
  a doublet would have to be quite heavy, with masses well beyond the
  range we consider here; $\phi$ can therefore not itself be part of
  such a doublet.  In principle $\phi$ can also mix with the SM Higgs
  $\Phi$; however, the resulting (renormalizable) couplings to
  electrons and muons would be uninterestingly small.} Most
lepton--specific scalar mediator models considered in the literature
assume the effective scalar couplings $c_l$ to be proportional to the
charged lepton masses, in which case the $e'_l$ follow the lepton mass
hierarchy.

\begin{figure}[H]
  \centering
    \includegraphics[page=2]{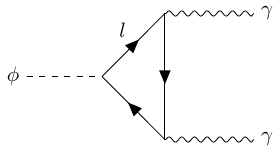}
    \caption{One loop contribution to the $\phi$-photon coupling.}
\label{fg:PhotonCoupling}
\end{figure}

A nonvanishing $e'_l$ leads at $1-$loop to an effective coupling of
the scalar to two photons, through the diagram shown in figure
\ref{fg:PhotonCoupling}. If $m_\phi < 2 m_l$, this decay can be used
to search for $\phi$ in the diphoton invariant mass distribution. The
decay width to photons is\footnote{The corresponding loop diagram for
  the SM Higgs was first computed numerically in \cite{Ellis:1975ap}
  and analytically in \cite{Shifman:1979eb}.} \cite{Chen:2018vkr}
\begin{equation}
  \Gamma(\phi\rightarrow \gamma \gamma)= \frac{\alpha^2m_\phi^3}{256\pi^3}
  \left\lvert\sum_{l=e,\mu,\tau} \frac{e'_l}{m_l}F_{1/2}(x_l) \right\lvert^2\,,
\end{equation}
where $x_l = \frac{4m_l^2}{m_\phi^2}$ and and the loop function $F_{1/2}$ reads
\begin{equation}
F_{1/2}(x_l)= \begin{cases}
-2x_l\left[1+(1-x_l)\arcsin^2(x_l^{-1/2})\right]&x_l \geq 1\\
-2x_l\left[1-\frac{1-x_l}{4}\left(-i\pi +\log\frac{1+\sqrt{1-x_l}}
    {1-\sqrt{1-x_l}}\right)^2\right]&x_l<1
\end{cases} \,.
\end{equation}
For $x_l \gg 1$, the loop function $F_{1/2} \rightarrow -4/3$.

For $g_D > e_l'$ and $2m_\chi < m_\phi$ the scalar mediator decays
mostly invisibly to the dark sector. The corresponding decay width is
\begin{equation}
  \Gamma(\phi \rightarrow \bar{\chi}\chi) = g_D^2\frac{m_\phi}{8\pi}
  \left(1-\frac{4m_\chi^2}{m_\phi}\right)^{3/2} \,.
\end{equation}
The main avenue for collider experiments to constrain this model is
then missing energy searches. On the other hand, for
$2m_\chi > m_\phi > 2 m_l$ the mediator decays mostly visibly to
leptons it directly couples to, with decay width
\begin{equation} \label{eq:phidec_l}
  \Gamma(\phi \rightarrow \bar{l}l)=e_l'^2\frac{m_\phi}{8\pi}
  \left(1-\frac{4m_l^2}{m_\phi}\right)^{3/2}\,.
\end{equation}
We'll be interested in scenarios where $|e'_l| \gsim 10^{-3}$ for
$l = e$ or $\mu$. The decay width (\ref{eq:phidec_l}) then corresponds
to a lifetime
$\tau_\phi \lsim 1.5\cdot 10^{-14} (1 \ {\rm MeV}) / m_\phi$ seconds;
even accounting for a Lorentz boost, $\phi \rightarrow l^+ l^-$ decays
will then usually be ``prompt'' if the corresponding decay is
kinematically allowed.

A light scalar coupling to muons can also explain the $(g-2)_\mu$
results. Its contribution is given by \cite{LEVEILLE197863,
  Chen:2018vkr}:
\begin{equation}
  \Delta a_\mu^\phi = \frac{e_\mu'^2}{8\pi^2} \int_0^1 d z
  \frac{m_\mu^2(1+z)(1-z)^2}{m_\mu^2(1-z)^2+m_{\phi}^2z}\,.
\end{equation}
For $m_\phi^2\ll m_\mu^2$ this simplifies to
\begin{equation}
  \Delta a_\mu^\phi \approx \frac{3e_\mu'^2}{16 \pi^2}
  \approx 1.9\times 10^{-10}\left(\frac{e_\mu'}{10^{-4}}\right)^2 .
\end{equation}

\subsection{Co--SIMP}

Finally, we consider the Co--SIMP mechanism proposed by Smirnov et al.
\cite{Smirnov:2020zwf}. A real scalar particle $\chi$ with strong
self--interactions is assumed as dark matter. An interaction with the
Standard Model of the form $\chi \chi e \rightarrow \chi e$ is
introduced in order to dissipate entropy from the dark sector whilst a
$Z_3$ symmetry stabilizes $\chi$.

\begin{figure}[H]
  \centering
    \includegraphics[page=3]{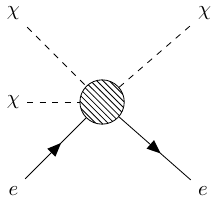}
    \caption{Effective operator of the Co-SIMP model we consider.}
\end{figure}

We consider an electrophilic version of this model, in which case the
relevant interaction is described by the effective operator
\begin{equation} \label{eq:simp_5}
\mathcal{O}_e = \bar{e}e \frac{\chi^3}{\Lambda^2}\,.
\end{equation}
Once again this does not respect the electroweak gauge symmetry; as
before gauge invariance can be restored by replacing the dim$-5$
operator of eq.(\ref{eq:simp_5}) by a dim$-6$ operator
$\propto \bar e \Phi e \chi^3 + h.c.$. Observation of cosmological
structures imply $m_\chi \gsim 5$ keV, while $m_\chi \lsim m_e$ is
required in order to avoid a WIMP--like freeze--out
\cite{Smirnov:2020zwf}. Note that this model does not contain
additional free parameters that allow to tune the relic density
independent of the laboratory limits which are the main topic of this
work. In order to compute the cosmologically preferred value of
$\Lambda$ we therefore solve numerically the Boltzmann equation
describing the freeze--out of $\chi$ \cite{Smirnov:2020zwf}:
\begin{equation}
  sH(T) x \frac{dY_\chi}{dx} = -s^3 \langle \sigma_{32} v^2\rangle
  \left( Y_\chi^2-Y_\chi Y_\chi^{\rm eq}\right)Y_e^{\rm eq}\,.
\end{equation}
Here $x = m_\chi/T$, $T$ being the temperature, $s$ is the total
entropy density, and $Y_\psi = n_\psi /s$ where $n_\psi$ is the number
density of $\psi$ particles, with $\psi \in \{\chi, e\}$ in our case;
the superscript ${\rm eq}$ denotes the equilibrium value of the
corresponding quantity. The thermally averaged cross section
\cite{PhysRevD.96.083521} is also obtained numerically from the
integral
\begin{equation} \label{eq:sigav}
  \langle \sigma_{32} v^2\rangle = \frac{1}
  {2n^{\rm eq}_e n^{\rm eq}_\chi n^{\rm eq}_\chi}
  \int \prod_{i=1}^5\frac{g_i d^3p_i}{(2\pi)^3 2E_i}(2\pi)^4
  \delta^4(p_1+p_2+p_3-p_4-p_5)f_1f_2f_3\overline{\lvert \mathcal{M} \rvert^2}\,.
\end{equation}
Here $g_i$ denotes the number of internal degrees of freedom of
particle $i$ ($1$ for $\chi$ and $4$ for $e$),
$f_i = 1 / (\exp{E_i/T}\pm 1)$ is the phase space distribution
function for the $i-$th particle in the initial state, and
$\overline{\lvert \mathcal{M} \rvert^2}$ is the averaged squared
matrix element for the relevant process
$\chi+\chi+e \rightarrow \chi + e$.

\section{Bounds from \texorpdfstring{$\mu^-$}{Muon} Decays}
\label{se:MuBounds}

\begin{figure}[ht]
\centering
\subfloat[]{
  \includegraphics[page=4]{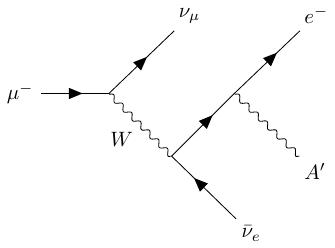}}%
\subfloat[]{
\includegraphics[page=5]{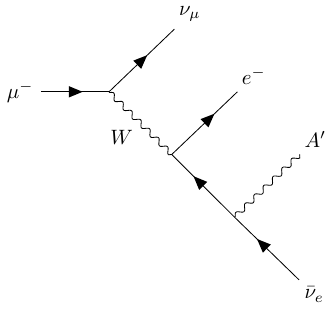}}
\caption{Diagrams contributing to $\mu^- \rightarrow e^- A' \nu_\mu \bar \nu_e$
  decays for the case that $A'$ couples to electron number.}
\label{fg:MuonBSMContribution}
\end{figure}

Having introduced the models we will consider, we turn to a discussion
of the decays which we use to derive bounds on the parameters of these
models. We begin with a discussion of the muon decay spectrum. Since
this has widely been used as a high precision test of the electroweak
theory \cite{TWIST:2011aa}, it might provide a good chance to
constrain our models. The double differential width for
$\mu^- \rightarrow e^- \nu_\mu \bar \nu_e$ decays at rest can be
written as \cite{ParticleDataGroup:2020ssz}:
\begin{equation} \label{eq:Diffrate}
  \frac{d^2\Gamma}{dxd\cos\theta} = \frac{m_\mu}{2\pi^3} W_{e\mu}^4 G_F^2
  \sqrt{x^2-x_0^2} \left[ F_{\text{IS}}(x) - P_\mu \cos\theta F_{\text{AS}}(x)
  \right]\,.
\end{equation}
Here $W_{e\mu} = (m_\mu^2+m_e^2)/2m_\mu$ is the maximum electron
energy (neglecting possible neutrino masses), $x = E_e / W_{e\mu}$ is
the rescaled electron energy, $x_0 = m_e/W_{e\mu}$ is its minimum
value, $P_\mu$ is the degree of muon polarization and $\theta$ is the
angle between the polarization vector of the muon and the outgoing
electron.  The functions for the isotropic part $F_\text{IS}(x)$ and
the anisotropic part $F_\text{AS}(x)$ are given by
\cite{ParticleDataGroup:2020ssz}:
\begin{align} \label{eq:F}
  F_\text{IS} &= x(1-x) + \frac{2}{9} \rho (4x^2-3x-x_0^2) + \eta x_0(1-x)\,;\\
  \nonumber
  F_\text{AS} &= \frac{1}{3} \xi \sqrt{x^2-x_0^2} \left[ 1 - x + \frac{2}{3}
                \delta \left( 4x - 4 + \sqrt{1-x_0^2} \right) \right]\,.
\end{align}
Here the Michel parameters $\rho,\eta,\xi$ and $\delta$ have Standard
Model values $3/4$, $0$, $1$ and $3/4$ respectively. Measurements of
these parameters have been used to put constraints on the effective
parameters of additional four fermions interactions. Here we use
precise measurements of the muon decay spectrum to put constraints on
four--body decays $\mu^- \rightarrow e^- \nu_\mu \bar \nu_e X$ where
$X$ is a light leptophilic mediator. The contributing Feynman diagrams
for the case of a vector mediator coupling to electron number are
shown in figure \ref{fg:MuonBSMContribution}. We assume that $X$ is
either long lived or decays invisibly, so that the four--body final
state has the same basic signature as the three--body final
state. However, the observable electron spectrum of this four--body
mode differs from the spectrum predicted by the SM.

The obvious Standard Model background is
the radiative muon decay,
$\mu^- \rightarrow e^- \gamma \nu_\mu \bar \nu_e$, where the photon
escapes detection but nevertheless carries some energy, thereby also
altering the electron spectrum.

Since our signal involves a four particle final state, this case
cannot be mapped directly onto eqs.(\ref{eq:Diffrate}) and
(\ref{eq:F}). However, when the dark mediator remains invisible the
observable final state has the same topology as in the SM, i.e. the
combination
\begin{equation} \label{eq:newmuwidth}
  \frac{d\Gamma(\mu\rightarrow e^- \bar{\nu_e} \nu_\mu)}{dx d\cos\theta} +
  e_l'^2 \frac{d\Gamma(\mu\rightarrow e^- \bar{\nu_e} \nu_\mu X)}
  {dx d\cos\theta}\big\rvert_{e_l'=1}
\end{equation}
will be observed. In order to derive estimates of possible constraints
we studied the sensitivity of fitting procedures similar to those
employed in the experimental determination of the spectral parameters.

The four body matrix elements were derived with help of the
\verb!Mathematica! package \verb!FeynCalc! \cite{Shtabovenko_2020}
followed by the numerical integration over all kinematic parameters
except $x$ and $\cos \theta$. The resulting two dimensional
distribution is then added to the pure Standard Model spectrum. The
Michel parameters that best describe this new distribution were
extracted from eqs.(\ref{eq:Diffrate}) and (\ref{eq:F}) by means of a
$\chi^2$ fit, using the same binning as in ref.~\cite{TWIST:2011aa}.
More exactly, we minimized
\begin{equation} \label{eq:chisq}
  \chi^2 = \sum_i \frac{\left[ \Gamma_i({\rm Michel}) - \Gamma_i(Z') \right]^2}
   {\Gamma_i(Z')} \,.
\end{equation}
Here $\Gamma_i$ is the muon decay width in the $i-$th bin;
$\Gamma_i({\rm Michel})$ is computed from eqs.(\ref{eq:Diffrate}) and
(\ref{eq:F}) and depends on the values of the Michel parameters, while
$\Gamma_i(Z')$ is computed from eq.(\ref{eq:newmuwidth}) and depends
on the mass and coupling of the new $Z'$ boson. Our final estimate for
the sensitivity of these measurements to the new coupling is obtained
by comparing these fitted Michel parameters to the measured values
\cite{ParticleDataGroup:2020ssz}:
\begin{equation} \label{eq:Michel}
  \rho = 0.74979 \pm 0.00026\, ; \ \ \eta = 0.057 \pm 0.034\,; \ \
  \delta = 0.75047 \pm 0.00034\, ; \ \ |P_\mu \xi| = 1.0009^{+0.0016}_{-0.0007}\,.
\end{equation}
We assume muon polarization $P_\mu = 1$, as predicted by the SM for
the relevant case of muons produced in meson decays.

A technical subtlety arises because experiments do not fit to the
whole spectrum: cuts have to be applied in order to cover detector
inefficiencies and blind spots. We modeled the effects of these cuts
by restricting our kinematical fit to the fiducial region covered by
the TWIST detector \cite{TWIST:2011aa}: approximately
$x \in [0.45, 0.98]$ and $\lvert \cos \theta \rvert \in
[0.54,0.96]$. We found that these cuts affect the sensitivity limit on
the coupling only by an $\mathcal{O}(1)$ factor, for the mediator
masses considered here. Our sensitivity limits should not be confused
with experimental bounds; we did not use real data, nor did we include
QED corrections when modeling the SM prediction for the decay
spectrum.  Our procedure should nevertheless give a reasonable
estimate of the sensitivity of the measurements of muon decays to the
new mediators.

We finally note that the measurement of the muon lifetime cannot be
used directly to constrain our models. In the SM this measurement is
used to determine the experimental value of $G_F$, which is a free
parameter of the theory. A deviation from the SM could therefore only
be detected by comparing this measurement with a second, independent
determination of $G_F$. Assuming unitarity of the quark mixing (CKM)
matrix, the experimental ``CKM unitarity test'' can be recast as a
measurement of $G_F$ -- with, however, much poorer precision
\cite{ParticleDataGroup:2020ssz}. Moreover, the emission of collinear
mediators can give rise to $\ln (m_\mu / m_X)$ enhanced terms in the
decay distribution, which cancel in the total muon decay width by the
Kinoshita--Lee--Nauenberg (KLN) theorem \cite{Kinoshita:1962ur,
  Lee:1964is} once loop diagrams are included. We therefore expect
that a comparison of different measurements of $G_F$ has much poorer
sensitivity to the light mediators we consider than the measurement of
the muon decay spectrum discussed above.

\section{Bounds from Leptonic Decays of Charged Pseudoscalar Mesons}
\label{se:BoundsPI}

At tree level the total width for the decay of a charged pseudoscalar
meson $P^\pm$ to a lepton pair is given by
\cite{Scherer:2002tk}
\begin{equation} \label{eq:Pdec}
  \Gamma(P^- \rightarrow l^-\bar\nu_l) = \frac{G_F^2 \lvert V_{q_1q_2}\rvert ^2}
  {4\pi} F_P^2 m_P m_l^2  \left(1-\frac{m_l^2}{m_P^2}\right)^2\,.
\end{equation}
Here $G_F$ is the Fermi constant, $m_P$ and $m_l$ are the meson and
lepton mass, respectively, $F_P$ is the $P$ decay constant, and
$V_{q_1q_2}$ is a CKM matrix element, $P^+$ being a $(q_1 \bar q_2)$
bound state. Owing to the $V-A$ nature of charged current weak
interactions, both the charged lepton and the neutrino ``like'' to be
left--handed, which however is forbidden by angular momentum
conservation. This leads to the well known helicity suppression
represented by the factor $m_l^2$ in eq.(\ref{eq:Pdec}). Evidently
this factor strongly suppresses the decay to an electron and
neutrino. This is often exploited for tests of lepton universality in
the ratio of decays to electrons and muons. The SM prediction for the
ratio of decay widths is
\begin{equation} \label{eq:Prat}
  R_P^{\rm SM} \equiv \frac{\Gamma(P\rightarrow e\nu(\gamma))}
  {\Gamma(P\rightarrow \mu\nu(\gamma))} =
  \left(\frac{m_e}{m_\mu}\right)^2 \left(
    \frac{m_P^2-m_e^2}{m_P^2-m_\mu^2}\right)^2 (1+\delta R_{\rm QED})\,,
\end{equation}
where $\delta R_{\rm QED}$ describes the effect of QED corrections,
including real photon emission. It is important to note that the
emission of a spin$-1$ boson does not change the helicity structure of
the amplitude, and therefore does not lift the helicity
suppression. In contrast, if a scalar or pseudoscalar particle couples
to the electron, the helicity suppression is removed, which can
enhance the electronic decay mode significantly. Since measurements
are in agreement with the SM prediction (\ref{eq:Prat}), this lifting
of the helicity suppression can be used to derive bounds on the
couplings of new light spin$-0$ particles.

In the limit of vanishing electron and neutrino masses, the total
width for $P \rightarrow e \nu_e \phi$ decays can easily be computed
analytically:\footnote{Our result agrees with that of
  ref.\cite{Berryman:2018ogk}, up to an overall factor of $2$ which
  arises because they consider a spin$-0$ particle coupling only to
  $\nu_L$. There is also a contribution $\propto e_e'^2 m_e^2$, which
  is IR divergent for $m_\phi \rightarrow 0$ \cite{Barger:1981vd}. As
  predicted by the KLN theorem, these terms are canceled by loop
  diagrams \cite{Pasquini:2015fjv}. Our simple expression
  (\ref{eq:Pdec3}) therefore accurately captures the most important
  contribution due to $\phi$ emission.}
\begin{equation} \label{eq:Pdec3}
\begin{split}
  \Gamma(P \rightarrow e \nu_e \phi) = \frac{e_e'^2 G_F^2 F_P^2 |V_{q_1q_2}|^2
    m_P^3} {384\pi^3} \Bigg[ 1 - \left(\frac{m_\phi}{m_P}\right)^6
&+ \left(\frac{m_\phi}{m_P}\right)^2\left(9+6\ln\left(\frac{m_\phi^2}{m_P^2}
  \right)\right)\\
&-\left( \frac{m_\phi}{m_P}\right)^4 \left(
  9 - 6\ln\left(\frac{m_\phi^2}{m_P^2}\right)\right) \Bigg].
\end{split}
\end{equation}
This expression manifestly avoids the $m_e^2$ suppression. In the
Co--SIMP model one instead has to emit three scalar $\chi$ particles,
leading to a considerably more complicated phase space integral;
however, since the new vertex again violates chirality, also in this
case the helicity suppression is lifted.

\begin{figure}[H]
\centering
\begin{minipage}{0.49\linewidth}
\includegraphics[width=\textwidth]{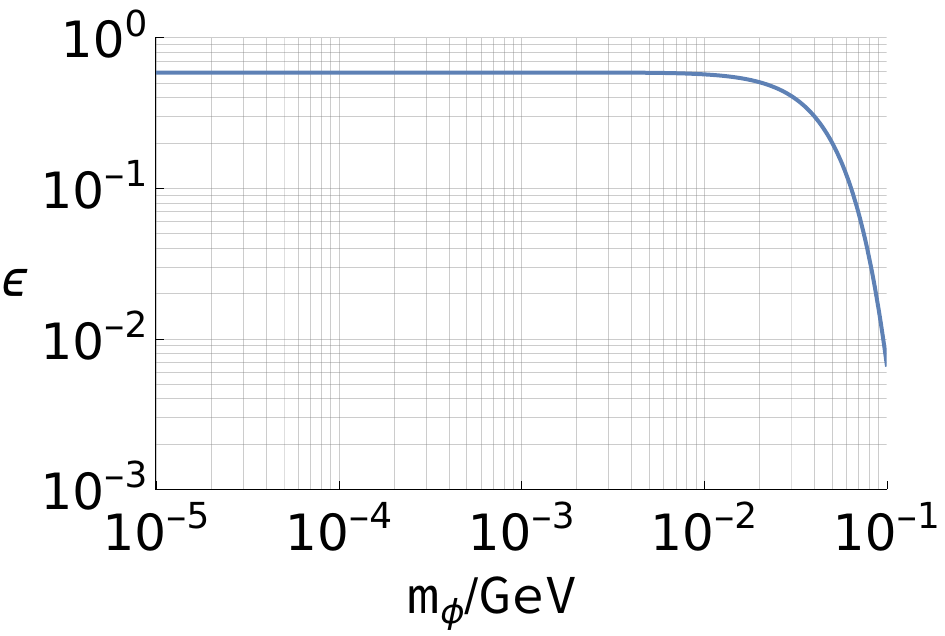}
\end{minipage}
\hfill
\begin{minipage}{0.49\linewidth}
\includegraphics[width=\textwidth]{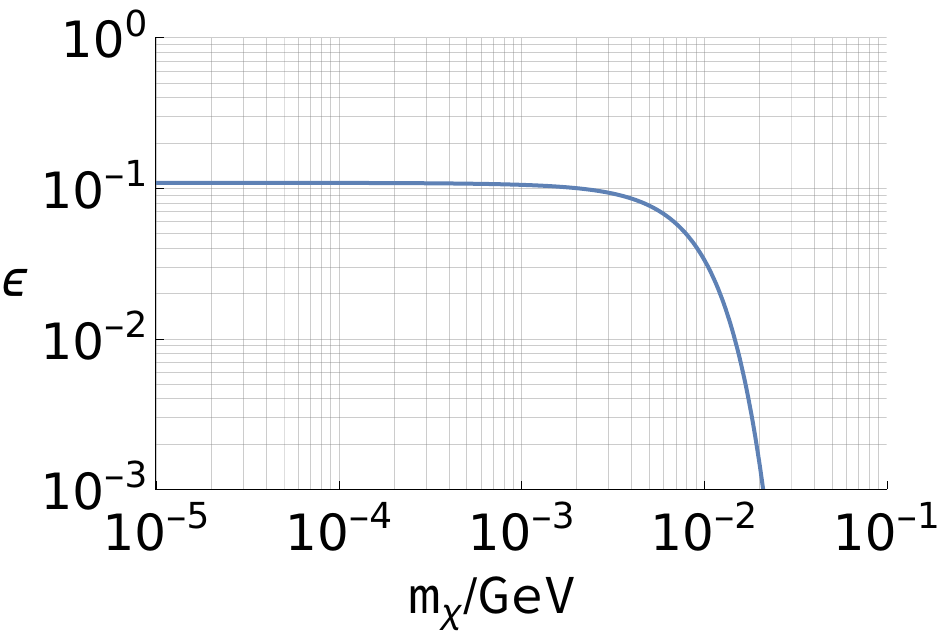}
\end{minipage}
\caption{Acceptance correction for $\pi^+ \rightarrow e^+ \nu_e \phi$ and $\pi^+ \rightarrow e^+\nu_e \chi \chi \chi$  decays due to the cut on $E_e$ by the PIENU experiment.}
\label{fg:PionAcceptance}
\end{figure}

Whenever the new scalars remain invisible the event will have the same
topology as a rare decay into $e+\nu_e$, albeit with a softer electron
energy spectrum. This change of the electron spectrum can reduce the
sensitivity due to kinematic cuts employed by the experiments. Bounds
are then extracted by saturating the maximal allowed difference
$\Delta_P$ between the theory prediction $R_p^{\rm SM}$ and the
experimental result $R_p^{\rm exp}$ , i.e.
\begin{equation} \label{eq:Pbound}
\Delta_P\geq\frac{\epsilon\Gamma(e^+\nu_e+ X)}{\Gamma(\mu\nu_\mu(\gamma))} \,.
\end{equation}
Here $X$ stands for either a single spin$-0$ mediator $\phi$ or for
the three Co--SIMP scalars $\chi$, and $\epsilon$ is an acceptance
correction factor due to the softer electron spectrum. For the most
precise measurement of $R_\pi$, by the PIENU collaboration
\cite{Aguilar-Arevalo:2017vlf}, this factor is computed as
follows. This experiment analyses decays of a stopped $\pi^+$ beam. In
order to discriminate between direct $\pi^+ \rightarrow e^+$ decays
and the dominant background from
$\pi^+ \rightarrow \mu^+ \rightarrow e^+$, an energy cut
$E_e \geq \SI{52}{\mega \eV}$ was used in the experimental
definition of $\pi^+ \rightarrow e^+ \nu_e$ decays, which have
a nominal positron energy of $E_e= \SI{69.8}{\mega \eV}$. If we want
to apply this analysis to our $\pi^+ \rightarrow e^+ \nu \phi$ decay,
the same cut on the positron energy should be applied. The resulting
acceptance correction is shown in figure \ref{fg:PionAcceptance}.

For the Co--SIMP model the acceptance is approximately constant in the
allowed mass range, with $\epsilon \approx 11\%$. This limitation does
not apply to $K^\pm$, $D^\pm$ and $D_s^\pm$ decays, which are studied
in flight, so that even the two--body decay mode has a broad energy
spectrum. The most precise measurement of $R_K$ comes from the NA62
collaboration \cite{thena62collaboration2013precision}. Here a
Kaon beam decays in flight inside the detector. Since the accepted
range of electron energies has a width of several $\SI{10}{\giga \eV}$
we simply assume that all decay modes have the same acceptance.

The decay of $D^\pm$ or $D_s^\pm$ mesons to $e \nu$ has not yet been
observed, but 90\%c.l. upper bounds on the corresponding branching
fractions have been set by the CLEO and Belle Collaboration
respectively \cite{CLEO:2008ffk, Belle:2013isi}. We use these to
derive limits on the couplings of our light spin$-0$ mediators.

\section{Results}
\label{se:Results}

Here we show our estimated sensitivity to the new coupling as function
of the mass of the postulated new boson, derived from muon and charged
meson decays. We also show existing limits found in the literature,
where we focus on the strongest bounds for the case at hand.

The relevant current bounds come from:
\begin{itemize}
\item \textbf{BaBar:} Dedicated search for the dark photon in
  $e^+ e^- \rightarrow \gamma A'; A' \rightarrow \text{invisible}$
  with CM energies near the $\Upsilon$ resonances. 90\% c.l. limits
  were derived on the dark photon coupling constant $\varepsilon^2$
  for $m_{A'}\leq \SI{8}{\giga \eV}$ \cite{Lees:2017lec}. This can
  directly be applied to our model with a vector mediator, with the
  replacement $e'_e = e \varepsilon$ \cite{essig2013dark}.
  
\item \textbf{CCFR:} Measurement of neutrino trident production events
  $\nu N \rightarrow \nu N \mu^- \mu^+$ using a muon-neutrino beam
  with average energy $\langle E_\nu \rangle = \SI{160}{\giga \eV}$.
  The observed $N_\text{CCFR}= 37.0 \pm 12.4$ events agree with the
  Standard Model prediction $N_\text{SM}=45.3 \pm 2.3$
  \cite{PhysRevLett.66.3117}. This is used to set limits on additional
  contributions from a vector mediator coupling to the muon-neutrino
  and to the muon \cite{Altmannshofer:2014pba}.
  
\item \textbf{Belle II}: Search for
  $e^+ e^- \rightarrow \mu^+ \mu^- Z'$ with beam energies of 4 and
  $\SI{7}{\giga \eV}$ where the $Z'$ is radiated off of one of the
  muons and decays invisibly with $m_{Z'}<\SI{6}{\giga \eV}$
  \cite{PhysRevLett.124.141801}. The limits can then be recast to
  muonphilic scalar mediators \cite{zhu2021probing}.

\item \textbf{PIENU:} Search for the three body decay
  $\pi^+ \rightarrow l^+\nu X$ where $l$ is an electron or muon and
  $X$ an invisible neutral boson. Pions were stopped in a detector and
  the spectrum of the charged lepton was measured. A search for the
  smooth signal spectrum was then carried out below the energy of the
  two body decay. This sets limits on the branching ratio
  $\Gamma\left(\pi^+\rightarrow e^+ \nu
    X\right)/\Gamma\left(\pi^+\rightarrow \mu^+ \nu\right)$ for
  $X-$masses in the range $0 < m_X < \SI{120}{\mega \eV}$
  \cite{Aguilar_Arevalo_2021}.

\end{itemize}

\begin{figure}[H]
\centering
\begin{minipage}{0.49\linewidth}
\includegraphics[width=\textwidth]{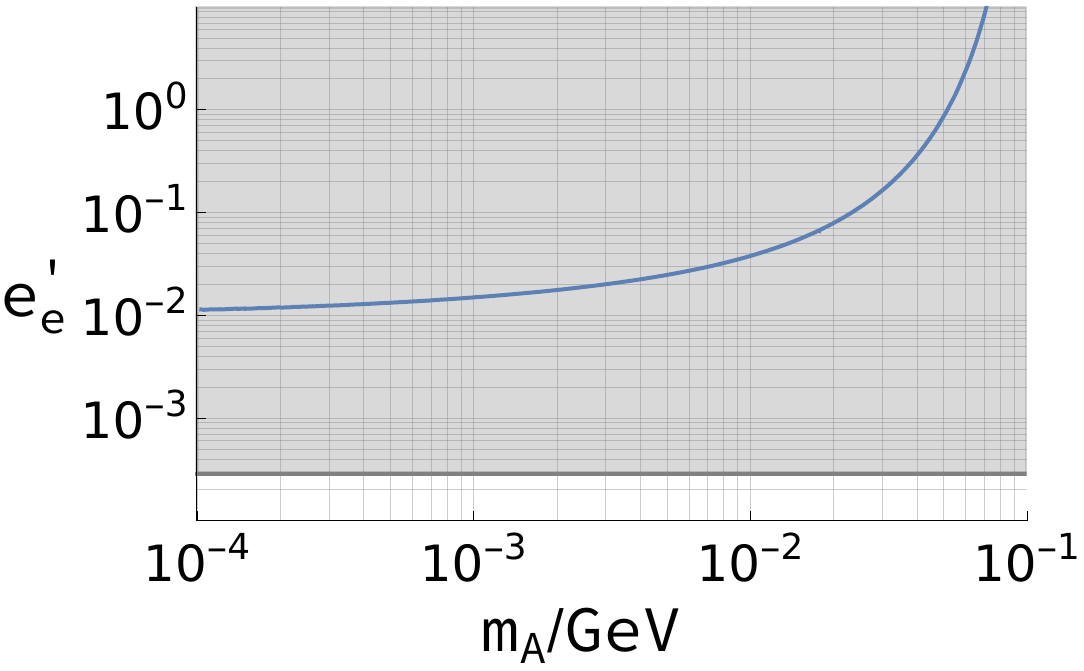}
\end{minipage}
\hfill
\begin{minipage}{0.49\linewidth}
\includegraphics[width=\textwidth]{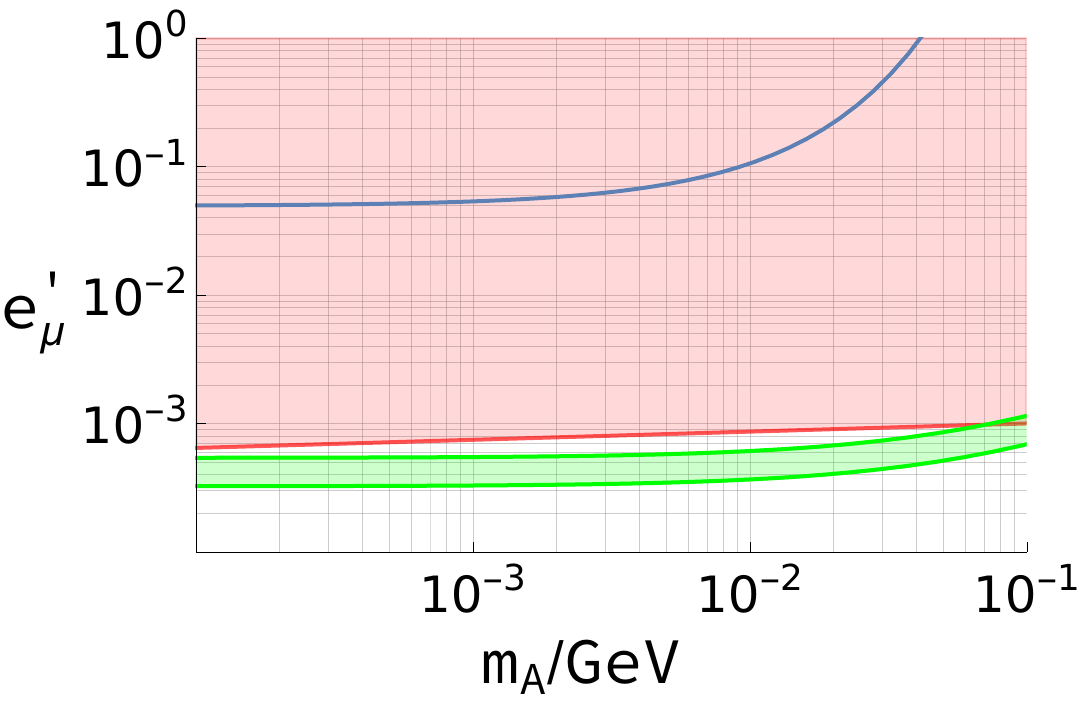}
\end{minipage}
\caption{The blue curves indicate our estimated 90\% c.l. sensitivity
  limit on the coupling constant between electron or muon and the
  spin$-1$ mediator $A'$ from existing measurements of muon decay.
  The grey shaded region in the left frame is excluded by the BaBar
  search for
  $e^+e^- \rightarrow \gamma A'(A'\rightarrow \text{invisible})$
  \cite{Lees:2017lec}, rescaled to $e'_e=\varepsilon e$, while the red
  shaded region in the right frame is excluded by an analysis
  \cite{Altmannshofer:2014pba} of CCFR data on
  $\nu N \rightarrow \nu N \mu^- \mu^+$. The green band in the right
  frame indicates parameters that bring the experimental and
  theoretical values of $(g-2)_\mu$ within $2\sigma$.}
\label{fg:MuPlot}
\end{figure}

Our estimated sensitivity of existing muon decay data to the couplings
of a new vector mediator are shown in Fig.~\ref{fg:MuPlot}.
Unfortunately these estimated sensitivities are considerably weaker
than the best existing bounds; the electron coupling is constrained by
\text{BaBar}, while the muon coupling is constrained by the
\text{CCFR} trident data. We find a considerably better sensitivity to
the electron coupling, since near--collinear $A'$ emission off the
electron, which is enhanced by a large logarithm, reduces the energy
of the electron and thus leads to an observable effect. In contrast,
near--collinear $A'$ emission off the muon neutrino leaves the electron
energy essentially unchanged. Below a mediator mass of
$\mathcal{O}(\SI{10}{\mega \eV})$ the bound on the Michel parameter
$\delta$ determines the sensitivity limit. For larger mediator masses
the sensitivity limit is set by the $\xi$ parameter. Of course, the
sensitivity is worse for larger mediator masses due to the closing
phase space.

\begin{figure}[H]
\centering
\begin{minipage}{0.49\linewidth}
\includegraphics[width=\textwidth]{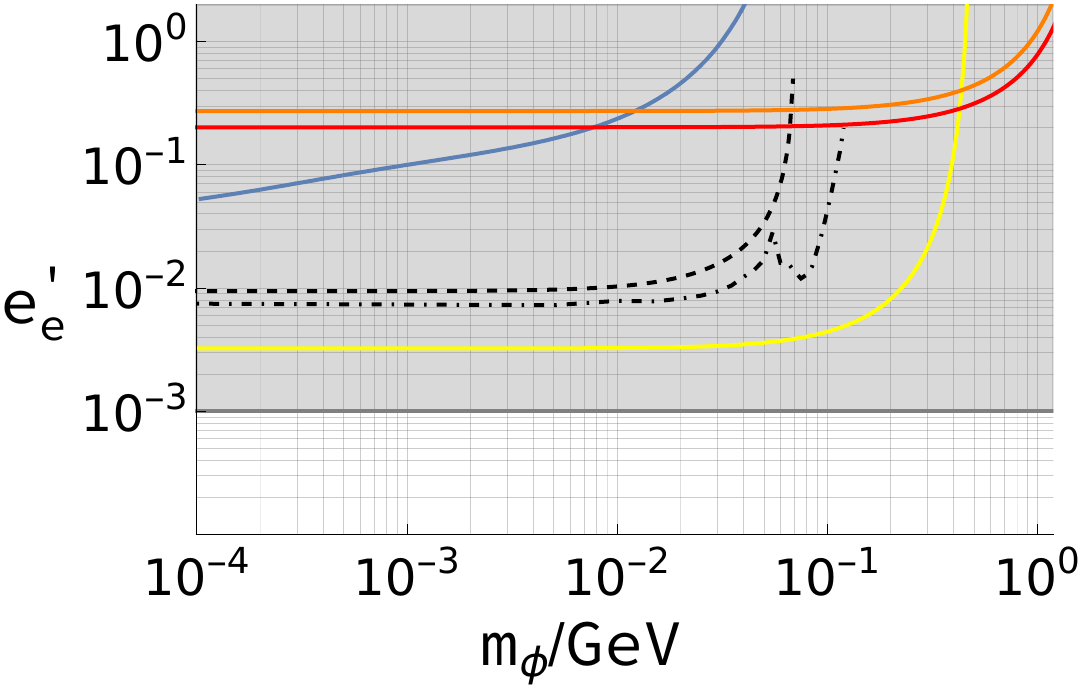}
\end{minipage}
\hfill
\begin{minipage}{0.49\linewidth}
\includegraphics[width=\textwidth]{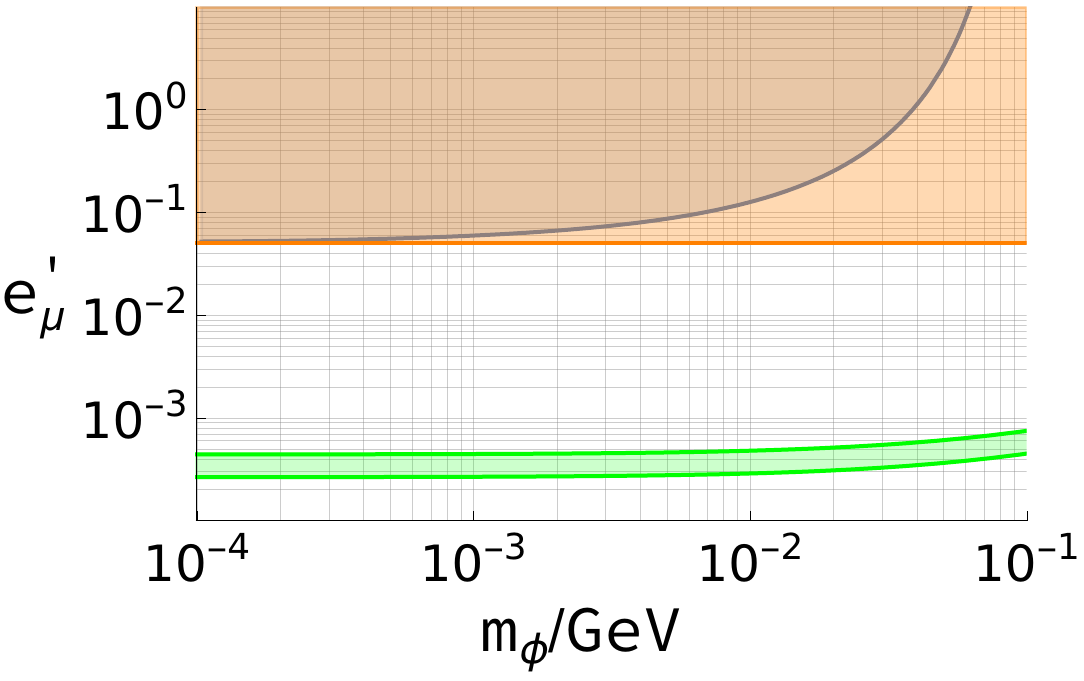}
\end{minipage}
\caption{90\% c.l. upper bounds on the coupling constant between
  electron or muon and the spin 0 mediator $\phi$. The blue curves in
  both frames indicate the sensitivity derived from muon decay. In the
  left frame, the dashed black, and solid yellow, red and orange lines
  indicate limits from additional contributions to
  $P^+ \rightarrow e^+\nu_e$ decays (dashed black: $P = \pi$; yellow:
  $P = K$; orange: $P = D$; red: $P = D_s$). The dot-dashed black
  line represents the limits derived from the result of the search of
  PIENU for $\pi^+\rightarrow e\nu X$; the drop of sensitivity at
  $m_\phi \sim \SI{55}{\mega \eV}$ is due to the signal being similar
  to $\pi^+ \rightarrow \mu^+\rightarrow e^+$
  \cite{Aguilar_Arevalo_2021}. The region shaded in gray in the left
  frame is excluded by an older \text{BaBar} limit recast to scalar
  mediators \cite{Essig:2013vha}. In the right frame the orange shaded
  region is excluded by a recast of a Belle II search for
  $e^+e^- \rightarrow \mu^+\mu^- A', \ (A'\rightarrow
  \text{invisible})$ \cite{zhu2021probing}, and the green band
  indicates parameters that bring the experimental and theoretical
  values of $(g-2)_\mu$ within $2\sigma$.}
\label{fg:ScPlot}
\end{figure}

Our projected bounds on the couplings of the scalar mediator $\phi$
are depicted in Fig.~\ref{fg:ScPlot}. Muon decay is much less
sensitive to these couplings than to those of a spin$-1$ mediator
shown in Fig.~\ref{fg:MuPlot}, since collinear emission of a soft
spin$-0$ boson off a fermion is suppressed. This also explains why the
bounds on $e'_e$ and $e'_\mu$ from muon decay are now quite similar.

However, bounds on $P^+ \rightarrow e^+ \nu_e$ decays, where $P^+$ is
a pseudoscalar ground state meson, lead to quite stringent bounds on
new scalar couplings of the electron. For $m_\phi \leq 400$ MeV the
strongest bound originates from Kaon decays; decays of $D^+$ and $D_s$
mesons are considerably weaker, but extend to larger mediator masses.
However, even the constraint from $K^+$ decays is somewhat weaker than
that from an older \text{BaBar} search for
$e^+ e^- \rightarrow \phi \gamma$ with invisible $\phi$. We do not
show bounds from charged meson decays on the muon coupling. Since the
helicity suppression of $P^+ \rightarrow \mu^+ \nu_\mu$ is much weaker
than for the electron mode, the resulting bounds on $e'_\mu$ are
considerably less stringent than those on $e'_e$, and are thus not
competitive with the constraint from the measurement of $g_\mu - 2$.

\begin{figure}[H]
 \centering
 \includegraphics[width=0.75\textwidth]{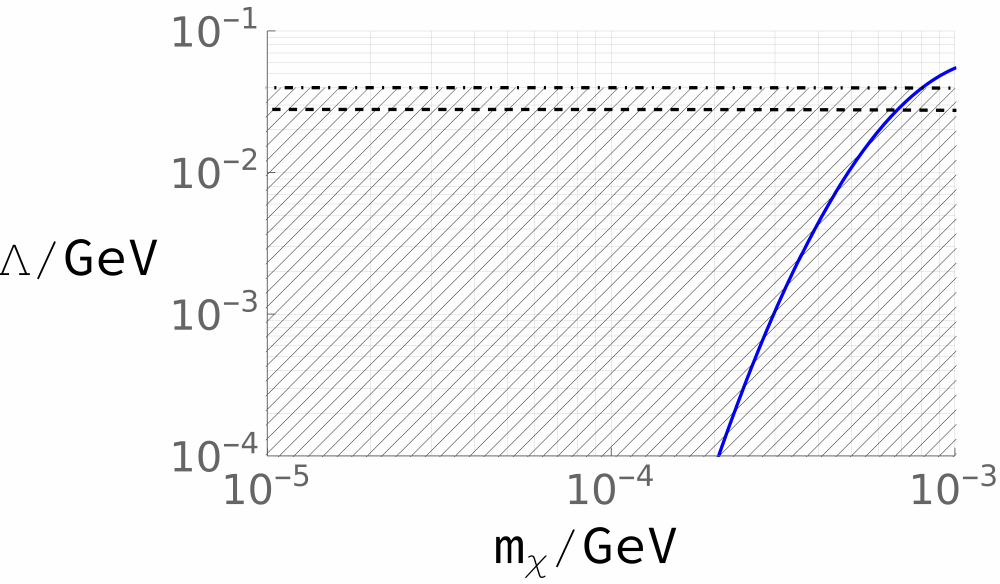}
 \caption{90\% c.l. lower bounds on the scale parameter $\Lambda$ of
   the Co--SIMP model. Again the black dashed and dot--dashed lines
   indicate limits from pion and kaon decays respectively. The solid
   blue line indicate parameter pairs that result in
   $\Omega_\chi h^2 \approx 0.12$ via the freeze--out mechanism in minimal
   cosmology.}
    \label{fg:CoSimpResults}
\end{figure}

Finally, constraints on the leptonic Co--SIMP scenario are summarized
in Fig.~\ref{fg:CoSimpResults}. The blue curve indicates the value of
$\Lambda$ required to obtain the correct relic density via
$\chi \chi e \rightarrow \chi e$ reactions, which change the number of
$\chi$ particles without changing the number of SM particles; this is
the essence of the Co--SIMP scenario of
ref.\cite{Smirnov:2020zwf}. The constraint from
$K^+ \rightarrow e^+ \nu_e$ decays excludes this value of $\Lambda$
unless $m_\chi > 0.8$ MeV. Note that for $m_\chi > 2 m_e / 3$ another
reaction becomes possible, $\chi \chi \chi \rightarrow e^+e^-$;
however, this is not a SIMP scenario any more. Meson decays therefore
exclude the whole region of parameter space where the correct $\chi$ relic
density is determined by the Co--SIMP reaction.\footnote{In our calculation
  we neglected the chemical potential of the electrons. Since the electron
  asymmetry is, like the baryon asymmetry, only ${\cal O}(10^{-9})$, this
  will affect the required value of $\Lambda$ only at very small $\chi$
  masses, where $\Lambda$ is already much below the bound from meson decays.}

\section{Conclusion}
\label{se:Conclusion}

In this paper we investigated the influence of leptophilic dark matter
models on two well measured decays of Standard Model particles:
deviations from the muon decay spectrum due to additional undetected
particles and the removal of the helicity suppression in the leptonic
decays of charged ground state pseudoscalar mesons. The measured muon
decay spectrum proved to be less sensitive to the new couplings than
published searches at $e^+e^-$ colliders and from neutrino trident
production. Limits on the electrophilic scalar coupling derived from
meson decays are an order of magnitude stronger than those from muon
decays but still weaker than existing limits by a factor of $\sim
2$. Hence a moderate precision improvement in leptonic kaon decays
could probe new regions of parameter space. On the other hand, the
measurement of the muon decay spectrum would have to become a lot more
accurate to yield competitive limits, which currently looks unlikely.
The muonphilic scalar limit at $m_\phi < \SI{1}{\mega \eV}$ is
competitive with the recast search from Belle II at
$e_\mu \sim 5\cdot 10^{-2}$, but lies well above the upper bound from
the measurement of the magnetic dipole moment of the muon.

For the purely electrophilic Co--SIMP we showed that the model is
ruled out for $m_\chi < \SI{800}{\kilo \eV}$ by the kaon decay
branching ratio because at allowed couplings $\chi$ is overproduced
by freeze--out. Above this mass the DM particle $\chi$ doesn't behave
like a SIMP any more, since its relic density is greatly affected by
annihilation into SM particles (namely $e^+e^-$ pairs). Our analysis
therefore excludes the electrophilic Co--SIMP scenario.

\bibliography{bibliography.bib}
\end{document}